\documentclass[a4paper,fleqn]{cas-dc}
\ExplSyntaxOn
\cs_gset:Npn \__first_footerline:
  { \group_begin: \small \sffamily \__short_authors: \group_end: }
\ExplSyntaxOff 
\usepackage{xcolor}

\usepackage{parskip}
\usepackage{mathrsfs}
\usepackage{amsfonts}
\usepackage{mathtools,cuted}
\usepackage[T1]{fontenc}
\usepackage[utf8]{inputenc}
\usepackage[english]{babel}

\usepackage{epstopdf}
 \usepackage{tabularx,ragged2e}
\newcolumntype{C}{>{\Centering\arraybackslash}X} 
\usepackage[numbers,sort&compress]{natbib}
\usepackage{hyperref}
\hypersetup{
    colorlinks=true,
    linkcolor=black,
    filecolor=magenta,      
    urlcolor=black,
    pdftitle={Overleaf Example},
    pdfpagemode=FullScreen,
    }

\usepackage{float}
\restylefloat{figure}

\begin{document}
\makeatletter
\@floatplacement
\makeatother

\let\WriteBookmarks\relax
\def\floatpagepagefraction{1}
\def\textpagefraction{.001}

\shorttitle{}
\shortauthors{Subrata Manna et~al.}

\title [mode = title]{Engineering Multi-wavelength Emission in All-Fiber Laser Mode-Locked Through Nonlinear Polarization Rotation}                      



%
\author[1]{Subrata Manna}[]






\affiliation[1]{organization={Ultrafast Fiber Optics \& Smart Photonic Technologies Lab, Department of Physics, IIT Hyderabad},
    addressline={Kandi}, 
    city={Telangana},
    postcode={502285}, 
    country={India}}


\author[1]{Amala Jose}[
   ]
    
\author[1]{K. Nithyanandan}[
   ]
\cormark[1]

\ead{nithyan@phy.iith.ac.in}

\begin{abstract}
The increasing demand for multi-wavelength optical sources to support dense wavelength-division multiplexing (DWDM) channels has driven the development of compact and reconfigurable multi-wavelength fiber lasers. Here, we demonstrate a continuously tunable and deterministically switchable multi-wavelength erbium-doped fiber laser based on nonlinear polarization rotation (NPR) in a compact all-fiber ring cavity. By controlling the intracavity birefringence, NPR acts as a reconfigurable comb filter that enables flexible wavelength selection without modifying the cavity architecture. The laser supports stable spectral states ranging from single- to seven-wavelength mode-locking, enabling reversible wavelength switching and activation/suppression of individual channels. The selectable spectral states can be mapped to binary bit operations, where each wavelength channel represents a controllable logical state. The behavior arises from the interplay between NPR-induced birefringent comb filtering and nonlinear phase modulation, providing a simple and compact platform for reconfigurable multi-channel ultrafast sources for DWDM and photonic signal processing.
\end{abstract}

\begin{keywords}
Mode-locked fiber laser \sep Multi-wavelength \sep Tunable \& Switchable laser \sep Nonlinear polarization rotation
\end{keywords}

\maketitle

\section{Introduction}
The continued growth of optical communications, multi-parameter fiber sensing, and photonic instrumentation has driven sustained interest in compact, reconfigurable multi-wavelength light sources. Remarkably, the balance of dispersion, nonlinearity, gain, and spectral filtering facilitates structured spectra and multi-wavelength emission in dissipative soliton mode-locked fiber lasers.~\cite{grelu2012dissipative}. Multi-wavelength fiber lasers (MWFLs) that emit several stable spectral lines from a single cavity are especially attractive for dense wavelength-division multiplexing (DWDM) \cite{DWDM}, multi-channel spectroscopy \cite{spectroscopy}, microwave or THz photonics \cite{microwave} and parallel sensing \cite{sensor} because they provide intrinsically synchronized channels, common timing, and simple packaging compared with independent lasers for each channel. However, producing stable, narrow-linewidth, and reconfigurable MWFLs in a single rare-earth-doped fiber cavity remains a challenge due to the intrinsic homogeneous gain broadening \cite{Homogeneous} and the resulting mode competition.\\


To overcome this limitation, researchers have pursued various strategies. One approach inserts spectral filters~\cite{boscolo2014pulse} or interferometric elements into the cavity—such as Fiber Bragg Gratings (FBGs)~\cite{FBG}, Mach-Zehnder interferometers~\cite{MZI,vasseur2004generation}, Sagnac interferometers~\cite{Sagnac}, and singlemode-multimode-singlemode (SMS) structures~\cite{SMS}—to create multiple low-loss spectral windows; however, these filter-based schemes increase cavity complexity and often constrain tunability. Another strategy exploits nonlinear inter-channel processes, including four-wave mixing (FWM)~\cite{FWM}, cross-phase modulation (XPM)~\cite{XPM}, and stimulated Brillouin scattering (SBS)~\cite{SBS}, to redistribute energy among wavelengths; while capable of generating numerous spectral lines, these methods typically require high pump power and strict phase-matching conditions. Yet another approach leverages material-based saturable-absorber (SA) technologies, such as SESAMs~\cite{SESAM}, CNTs~\cite{CNT}, graphene~\cite{Graphene}, and other 2D materials~\cite{2D}, to mention a few. Although effective for robust mode locking and self-starting, most of these material-based SA techniques suffer from limited spectral tunability, a low damage threshold, and long-term stability issues, depending on cavity architectures and environmental conditions.

Another category of SA utilizes interferometric or topology-dependent nonlinear loss mechanisms, such as nonlinear polarization rotation (NPR)~\cite{NPR}, nonlinear optical loop mirror (NOLM)~\cite{xin2022low}, and nonlinear amplifying loop mirror (NALM)~\cite{NALM}. Among these artificial SA-based methods, NPR has emerged as a robust mechanism for enabling multi-wavelength~\cite{MWL}, tunable~\cite{Tunable}, and switchable~\cite{switchable} operation. NPR is an all-fiber, low-loss technique~\cite{NPR_ML} that uses Kerr nonlinearity along with intracavity birefringence and polarization-dependent elements (polarizer or polarization-dependent isolator, polarization controllers) to produce an intensity-dependent transmission. 
For instance, in 2006 Feng et. al. reported an NPR-based ring fiber laser~\cite{Feng} employing a Fabry-Pérot thin-film filter that produced multi-wavelength emission, while Zhang et. al. achieved as many as 25 lines using a Lyot filter~\cite{Zhang}; however, both systems operated in the continuous-wave (CW) regime. Although effective, filter-based approaches inevitably add complexity and limit flexibility. In contrast, Gong et. al. demonstrated a dual-wavelength NPR-based mode-locked fiber laser, albeit not in a fully all-fiber configuration~\cite{Gong}. Later, Xu et. al. investigated an all-normal-dispersion Yb-doped fiber laser with tri-wavelength dissipative soliton operation, achieving fixed 16.4 nm spacing between adjacent lines~\cite{Xu}. Similarly, a dual- and tri-wavelength mode-locked Er-doped fiber laser was reported with switchable spectral intervals of 12.67, 33.32, and 43.4 nm by carefully tuning polarization controllers (PCs)~\cite{Song}. 

In the 2 $\mu$m regime, Yan et. al. demonstrate a tunable and switchable dual-wavelength ultrafast thulium-doped fiber laser, achieving dual-wavelength mode-locking with continuous tunability across a broad spectral range of 1852-1886 nm~\cite{Yan}. Hybrid approaches have also been explored to scale the number of wavelengths and improve stability.
Luo et al. (2011) reported a hybrid NPR~\cite{Luo} SESAM cavity, incorporating 1.6 m of polarization-maintaining fiber (PMF) with high birefringence. This design compressed the channel spacing and enabled up to seven stable mode-locked wavelengths in the communication band. 

Furthermore, in conventional multi-wavelength mode-locked lasers, pulses at different wavelengths generally experience distinct group velocities due to cavity dispersion, leading to asynchronous pulse trains that circulate independently and periodically collide inside the cavity~\cite{hamdi2022superlocalization, wu2023synchronization}. Achieving temporal synchronization, therefore, requires precise dispersion engineering or active group-delay compensation, as demonstrated via intracavity group-delay modulation~\cite{mao2021synchronized}. Such approaches, while effective, typically rely on programmable pulse shapers or additional dispersion-control elements, increasing system complexity.

\begin{figure}[H]
    \centering
    \includegraphics[width=\linewidth]{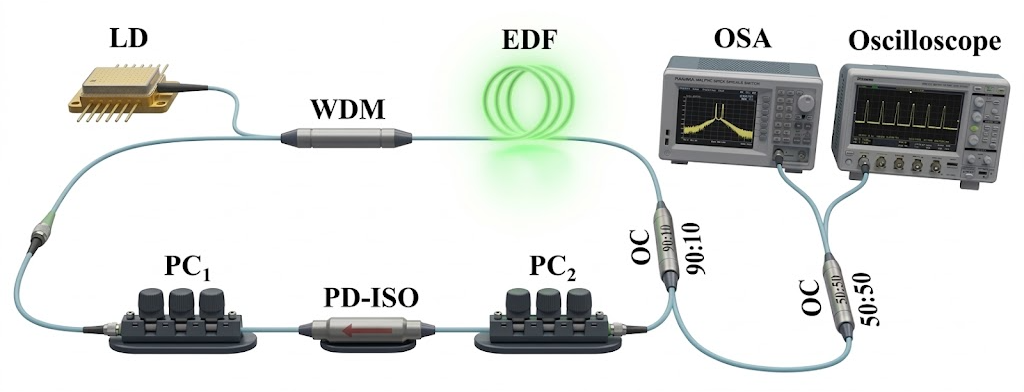} 
    \caption{Experimental setup of the NPR-based mode-locked erbium-doped fiber laser (EDFL). WDM: Wavelength Division Multiplexing, EDF: Erbium Doped Fiber, SMF: Single Mode Fiber, OC: Output Coupler, PC: Polarization Controller, PD-ISO: Polarization Dependent Isolator, LD: Laser Diode.}
    \label{Cavity}
\end{figure}


Despite significant progress, no fully alignment-free, purely NPR-based all-fiber erbium-doped fiber laser has yet demonstrated stable mode locking with up to seven wavelength channels. Moreover, such systems have not simultaneously achieved the following functions: (i) extended single wavelength tunability, (ii) collective multi-wavelength tuning with partially preserved spectral spacing, and (iii) deterministic, reversible, and combinational switching within a single cavity architecture.

In this work, we experimentally demonstrate an erbium-doped fiber laser (EDFL) passively mode-locked via NPR, incorporating (i) precise intracavity birefringence engineering, (ii) flexible polarization control, and (iii) optimized pump-power management within a fully all-fiber architecture. The synergistic interaction of these mechanisms enables a stable, highly reconfigurable, and dynamically controllable multi-wavelength source exhibiting both extended tunability and deterministic switchability. Moreover, in many previously reported systems, different wavelengths evolve as independent pulse trains. In contrast, we observe a single composite pulse envelope comprising phase-locked spectral components, confirming intrinsic temporal synchronization operation. Beyond high-order multi-wavelength generation, the laser provides continuous single wavelength tunability over an extended spectral range. In dual- to quadruple-wavelength regimes, the spectral peaks translate collectively while maintaining nearly constant inter-peak spacing, indicating synchronized intracavity dynamics. Furthermore, the system enables deterministic and reversible spectral switching, allowing selective activation of individual channels and arbitrary subset combinations within multi-wavelength states—without altering the cavity configuration.

\section{Experimental setup}
The schematic representation of the experimental setup is depicted in Fig.~\ref{Cavity}, illustrating a mode-locked EDFL based on NPR. 
At the core of the setup is a 3-meter-long Erbium-doped fiber (EDF), which serves as the gain medium. The EDF is optically pumped by a 976 nm laser diode capable of delivering up to 1W of power, ensuring sufficient population inversion for stimulated emission. The pump light is coupled into the fiber through a wavelength division multiplexer (WDM).

To maintain stable unidirectional pulse propagation while precisely controlling polarization states, the laser cavity incorporates two polarization controllers (PCs) and a polarization-dependent isolator (PD-ISO). These elements play a crucial role in sustaining mode-locking through NPR by adjusting the polarization-dependent transmission of the saturable absorber effect. The laser output is extracted from the cavity via a 10\% port of a 10:90 output coupler (OC), allowing most of the intracavity power to circulate while a fraction is utilized for analysis.

The total cavity length is extended to 25 meters by incorporating a 17-meter section of single-mode fiber (SMF) to enhance the nonlinear effect/ interaction associated with Kerr nonlinearity. This configuration results in an overall anomalous dispersion of approximately -0.17 ps², which is conducive to the formation of solitonic pulses. 

A combination of high-precision characterization techniques is employed to study pulse evolution and analyze spectral properties in real time. The optical spectrum of the output pulses is measured using a Yokogawa AQ6370D optical spectrum analyzer, which offers a resolution of 0.02 nm. The temporal characteristics and pulse dynamics are observed with a 4 GHz Tektronix MSO64 oscilloscope, while a 5 GHz Thorlabs DET08CFC/M photodetector is used to convert optical signals into electrical signals for further analysis. These measurement tools ensure a comprehensive characterization of the laser's performance, allowing for an in-depth understanding of soliton formation and propagation dynamics.

\section{Working Principle}
\begin{figure}[htb]
    \centering
    \includegraphics[trim={0.2cm 0.2cm 0cm 0.2cm},clip,width=\linewidth]{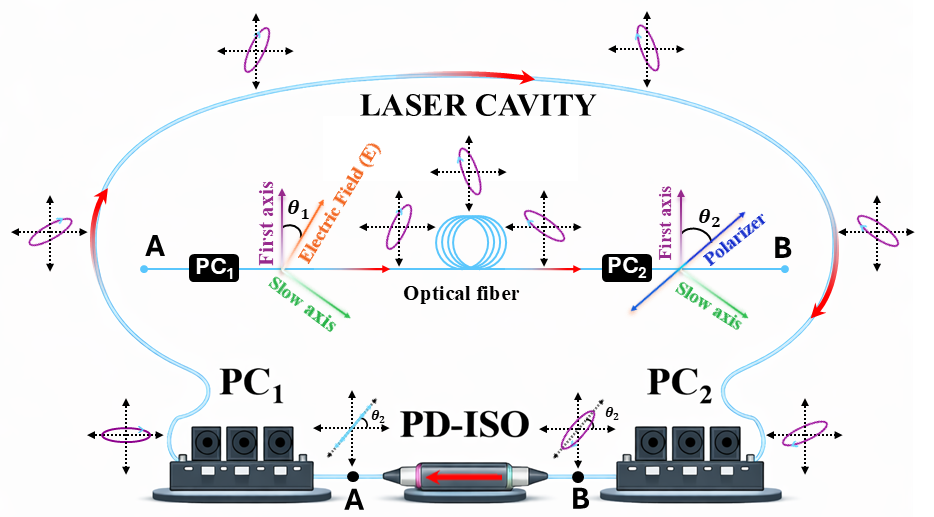} 
    \caption{Polarization states of mode-locked laser}
    \label{Cavity2}
\end{figure}
The operating principle of the NPR-based mode-locked fiber laser is governed by intensity-dependent polarization evolution arising from the combined effects of fiber birefringence and Kerr nonlinearity, followed by polarization-selective transmission. 
The NPR mechanism originates from the interference of two orthogonally polarized field components that propagate along the birefringent axes of the fiber. These components accumulate different linear birefringent and nonlinear Kerr-induced phase shifts during propagation. Upon projection by the polarization-selective element, the phase difference between the orthogonal components is converted into an intensity-dependent transmission, which effectively mimics the behavior of a fast saturable absorber and enables passive mode locking. The laser cavity can be effectively modeled as a birefringent fiber segment enclosed by two polarization controllers (PCs), with a polarization-dependent isolator (PD-ISO) placed between them. The PD-ISO simultaneously functions as a polarizer and an analyzer, providing both polarization discrimination and unidirectional operation. The transmission behaviour of the laser cavity is conceptually equivalent to the system depicted in Fig. \ref{Cavity2}. As the laser field passes through the PD-ISO, it emerges in a well-defined linear polarization state. The polarization controller PC-1 subsequently converts the linear polarization into a left/right-handed elliptical/circular polarization state. This polarized light can be expressed as a superposition of two orthogonal polarization components, allowing the intracavity electric field to be decomposed along the principal birefringent axes x and y of the fiber as
\begin{equation}
\mathbf{E}_{\mathrm{in}} = E_0
\begin{pmatrix}
\cos \theta_1 \\
\sin \theta_1
\end{pmatrix},
\label{eq:Ein}
\end{equation}
where $E_0$ is the field amplitude and $\theta_1$ denotes the polarization angle between the polarization direction of the light and the fiber fast axis.
As the optical field propagates through a fiber of length $L$, the combined effects of self-phase modulation (SPM) and cross-phase modulation (XPM) introduce an intensity-dependent phase shift.
The relative phase difference between the two components consists of two distinct contributions,
\begin{equation}
\Delta \Phi = \Delta \Phi_{\mathrm{L}} + \Delta \Phi_{\mathrm{NL}} ,
\label{eq:total_phase}
\end{equation}

where $\Delta \Phi_{\mathrm{L}}$ arises from linear birefringence and $\Delta \Phi_{\mathrm{NL}}$ originates from nonlinear Kerr effects. These nonlinear effects effectively induce different plane of polarization rotation states. The extent of this polarization rotation is directly proportional to the intensity of the circulating light.  
After propagation through the birefringent fiber, the electric field is expressed as
\begin{equation}
\mathbf{E}_{\mathrm{fiber}} = E_0
\begin{pmatrix}
\cos \theta_1 \, e^{i \phi_x} \\
\sin \theta_1 \, e^{i \phi_y}
\end{pmatrix}.
\label{eq:Efiber}
\end{equation}
where $\phi_x$ and $\phi_y$ are the accumulated propagation phases of the two orthogonally polarized components. At the final stage, the PD-ISO selectively transmits specific polarization states that align with the axis of the polarizer. The analyzer transmission axis is represented by a unit vector
\begin{equation}
\hat{\mathbf{e}}_{\theta_2} =
\begin{pmatrix}
\cos \theta_2 \\
\sin \theta_2
\end{pmatrix}.
\label{eq:analyzer_axis}
\end{equation}

The transmitted field is obtained by projecting the fiber output field onto this axis:

\begin{equation}
E_{\mathrm{out}} = \hat{\mathbf{e}}_{\theta_2}^{\,T} \cdot \mathbf{E}_{\mathrm{fiber}} .
\label{eq:Eout}
\end{equation}

Here, $\theta_2$ denotes the angle between the transmission axis of the analyzer (PD-ISO combined with PC-2) and the fast birefringent axes of the optical fiber.
 It defines the polarization direction selected and transmitted by the analyzer. So, the normalized transmission is defined as
\begin{equation}
T = \frac{\lvert E_{\mathrm{out}} \rvert^{2}}{E_0^{2}} .
\label{eq:transmission}
\end{equation}
\begin{align}
T &= \cos^2 \theta_1 \cos^2 \theta_2 + \sin^2 \theta_1 \sin^2 \theta_2   + \frac{1}{2} \sin 2\theta_1 \sin 2\theta_2 \notag \\
&\quad \times \cos ({\Delta\Phi_{L}} + {\Delta\Phi_{NL}})
\end{align}
where,\begin{equation}
\Delta\Phi_{L} = \frac{2\pi L}{\lambda} (n_y - n_x)
\end{equation}
\begin{equation}
\Delta\Phi_{NL}= -\frac{2\pi L n_2 P}{3 A_{\text{eff}} \lambda} \cos 2\theta_1
\end{equation}
 The parameters $n_{x}$ and $n_{y} $correspond to the refractive indices of the fiber's fast and slow axes, respectively. The operating wavelength is represented by $\lambda$, while $n_{2}$ signifies the nonlinear Kerr coefficient of the fiber. The instantaneous peak power of the input signal is given by $P$, and $A_{eff}$ denotes the effective mode area of the fiber core. In essence, the coordinated operation of PC-1, PD-ISO, and PC-2 forms a polarization-selective, intensity-dependent system that enables stable multi-wavelength operation and passive mode-locking. The NPR mechanism exhibits intrinsic wavelength-dependent and inhomogeneous intensity-dependent transmission characteristics, which reduce mode competition arising from the homogeneous gain broadening of the erbium-doped fiber. In addition, the configuration simultaneously functions as an all-fiber Lyot birefringence filter. Owing to the controlled birefringent phase delay accumulated along the SMF and the polarization-maintaining fiber segments of the PD-ISO, the system exhibits periodic spectral transmission maxima, thereby operating as an effective comb filter\cite{xu2008two,kobtsev2025birefringent}. This dual functionality enables simultaneous intensity modulation and spectral filtering within a unified, reconfigurable, and fully fiber-integrated architecture.

\begin{figure}[htb]
    \centering
\includegraphics[width=\linewidth]{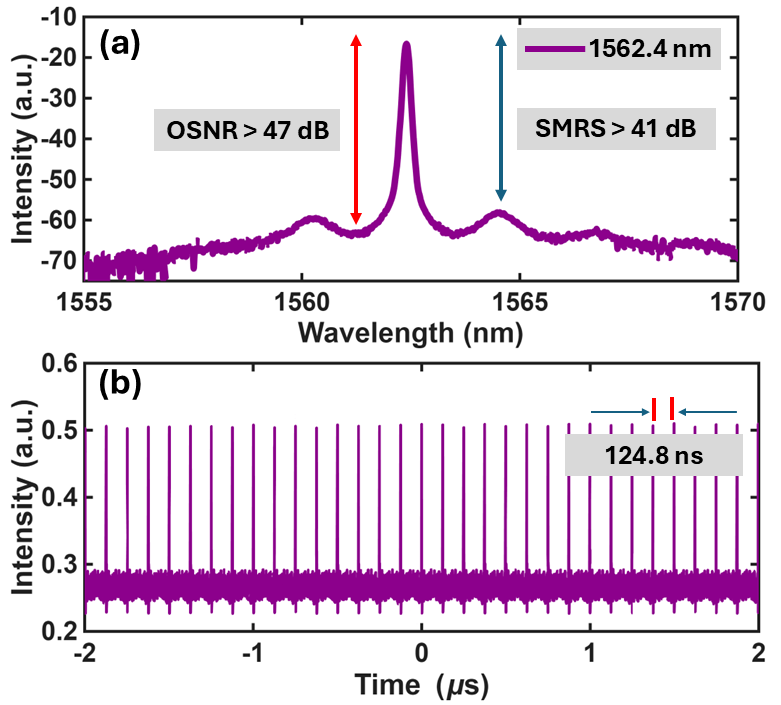} 
    \caption{ (a) Single wavelength NPR-based mode-locked Output spectrum and (b) the corresponding pulse train.}
    \label{single}
\end{figure}
\section{Results and discussion}
The measured pump power required to initiate mode-locking is approximately 18 mW, which is substantially lower than values typically reported for comparable fiber laser systems. This low threshold reflects the high efficiency of the cavity design and the effective polarization control enabled by the NPR mechanism. As the pump power is gradually increased beyond this threshold, the laser can be driven into a multi-wavelength mode-locked regime under suitable polarization conditions.
 
\begin{figure*}
    \centering
    \includegraphics[width=\linewidth]{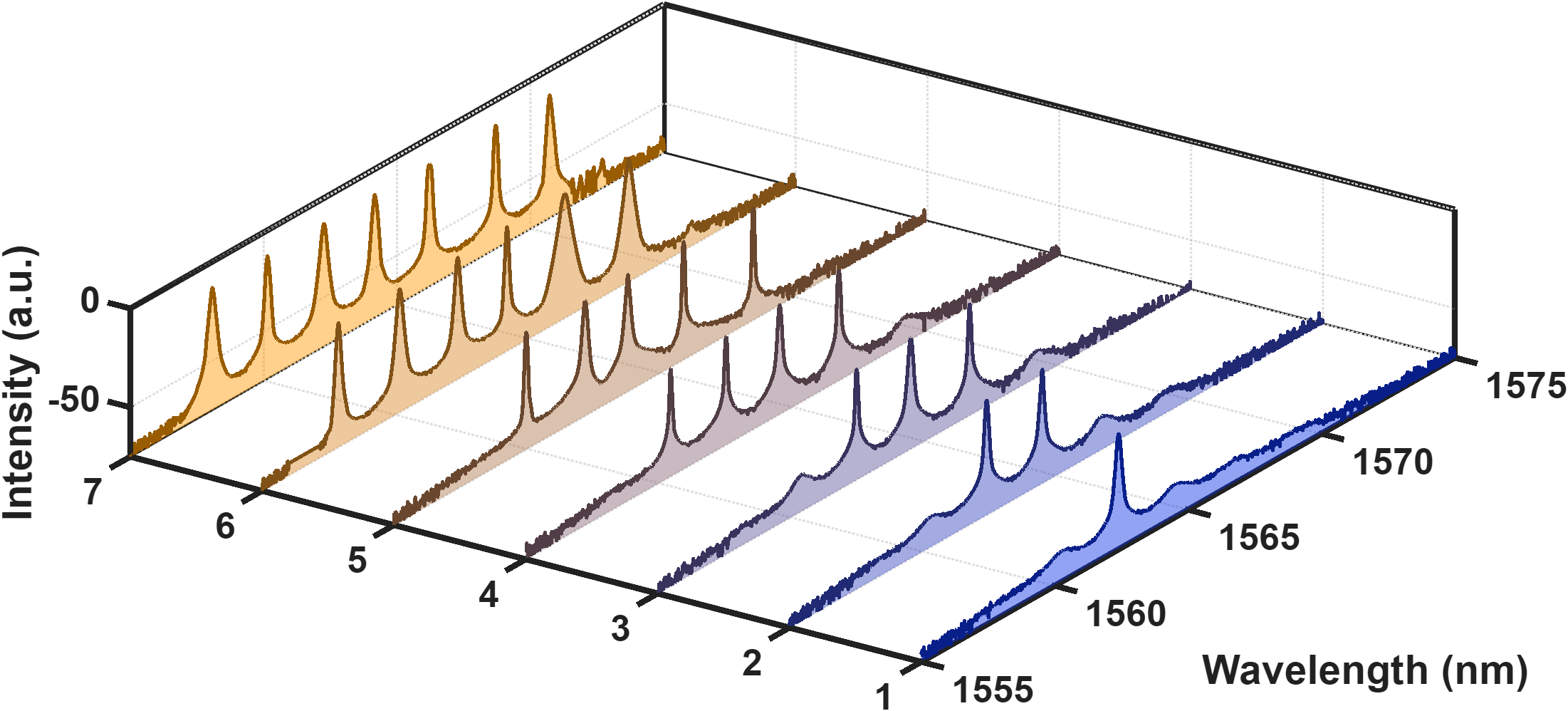} 
    \caption{Output spectra corresponding to  multi-wavelength operation from one to seven spectral peaks.}
    \label{7wl waterfall}
\end{figure*}
For specific polarization settings, the cavity favors stable single-wavelength mode-locked operation. In this regime, only one wavelength satisfies the NPR-induced mode-locking condition, resulting in the emergence of a single dominant spectral peak in the optical spectrum, as shown in Fig.~\ref{single}(a). No residual CW components or secondary lasing lines are detected across the spectral bandwidth, indicating complete suppression of competing longitudinal modes. The dominant lasing peak is centered at 1562.40 nm, and the spectrum exhibits a high optical signal-to-noise ratio (OSNR) exceeding 47 dB, confirming a low amplified spontaneous emission (ASE) background and excellent spectral purity. In addition, the side-mode suppression ratio (SMSR) exceeds 41 dB, demonstrating strong mode discrimination within the cavity. The emission linewidth is narrow, with a measured 3-dB spectral bandwidth of approximately 0.1 nm, indicative of strong phase coherence. The smooth spectral envelope and the absence of Kelly sidebands further show that the laser operates in a stable mode-locked dissipative soliton regime with minimal amplitude fluctuations. The temporal characteristics of the single-wavelength output further corroborate stable mode-locked operation. The measured time-domain intensity profile, shown in Fig.~\ref{single}(b), reveals a highly uniform pulse train extending over the full observation window. The pulse-to-pulse separation is approximately 124.8 ns, which corresponds directly to the cavity round-trip time and yields a repetition rate of 8.014 MHz, consistent with the cavity length. 
\subsection{Multi-wavelength Mode locked fiber laser}
By precise adjustment of the intracavity polarization controllers, the wavelength-dependent loss of the NPR-based fiber ring cavity can be effectively engineered.  The NPR-based cavity operates as a dual-functional element, as mentioned earlier. It provides ultrafast saturable absorption through the intensity-dependent polarization rotation~\cite{NPR_SA}. Also, the accumulated linear birefringence gives rise to a periodic transmission spectrum ~\cite{NPR_filter}, imposing multiple discrete spectral passbands within the erbium gain envelope. The coexistence of these two functions fundamentally alters the intracavity gain-loss balance: the wavelength-dependent loss partially lifts the homogeneous gain constraint, while the nonlinear loss promotes power equalization among competing modes. As a result, stable multiwavelength mode locking can be sustained within a single homogeneously broadened gain medium without the need for additional spectral filtering elements. This polarization-controlled loss modulation plays a decisive role in selecting the lasing wavelengths, determining the number of simultaneously oscillating modes, and ensuring spectral uniformity among the multiple wavelengths. As a result, the laser exhibits exceptional flexibility, supporting stable mode-locked operation across different multi-wavelength regimes.
\begin{figure}[htb]
    \centering
    \includegraphics[width=0.8\linewidth]{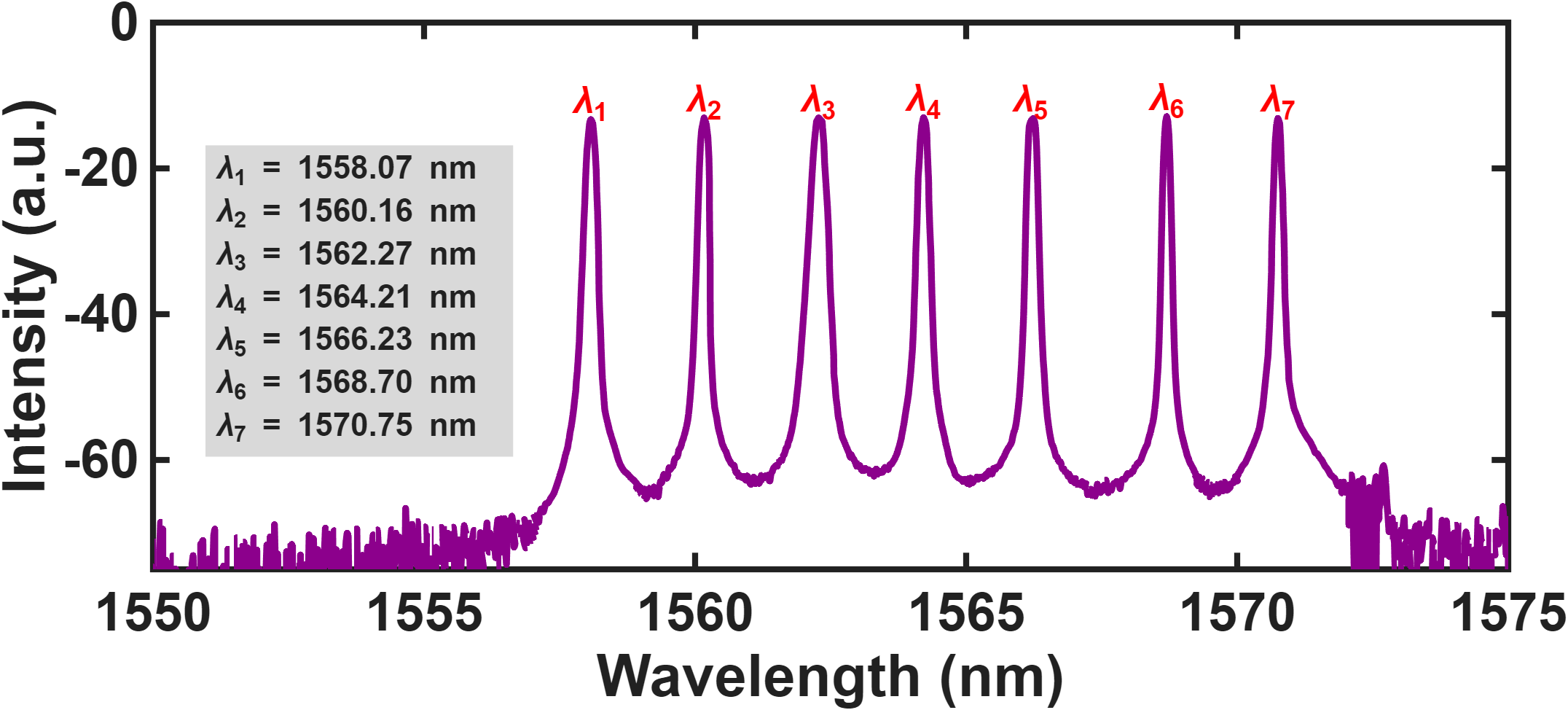} 
    \caption{Measured seven-wavelength output spectrum of the mode-locked fiber laser.}
    \label{7 wl}
\end{figure}
Our experimental observations reveal that the fiber laser is capable of sustaining not only single- and dual-wavelength mode-locking, but also higher-order multi-wavelength states, up to seven-wavelength mode-locked operation, as illustrated in Fig.~\ref{7wl waterfall}. These distinct operating regimes are accessed by fine-tuning the polarization controllers within a controlled angular range and the pump power, without any modification to the cavity architecture.
\begin{figure}[htb]
    \centering
    \includegraphics[width=\linewidth]{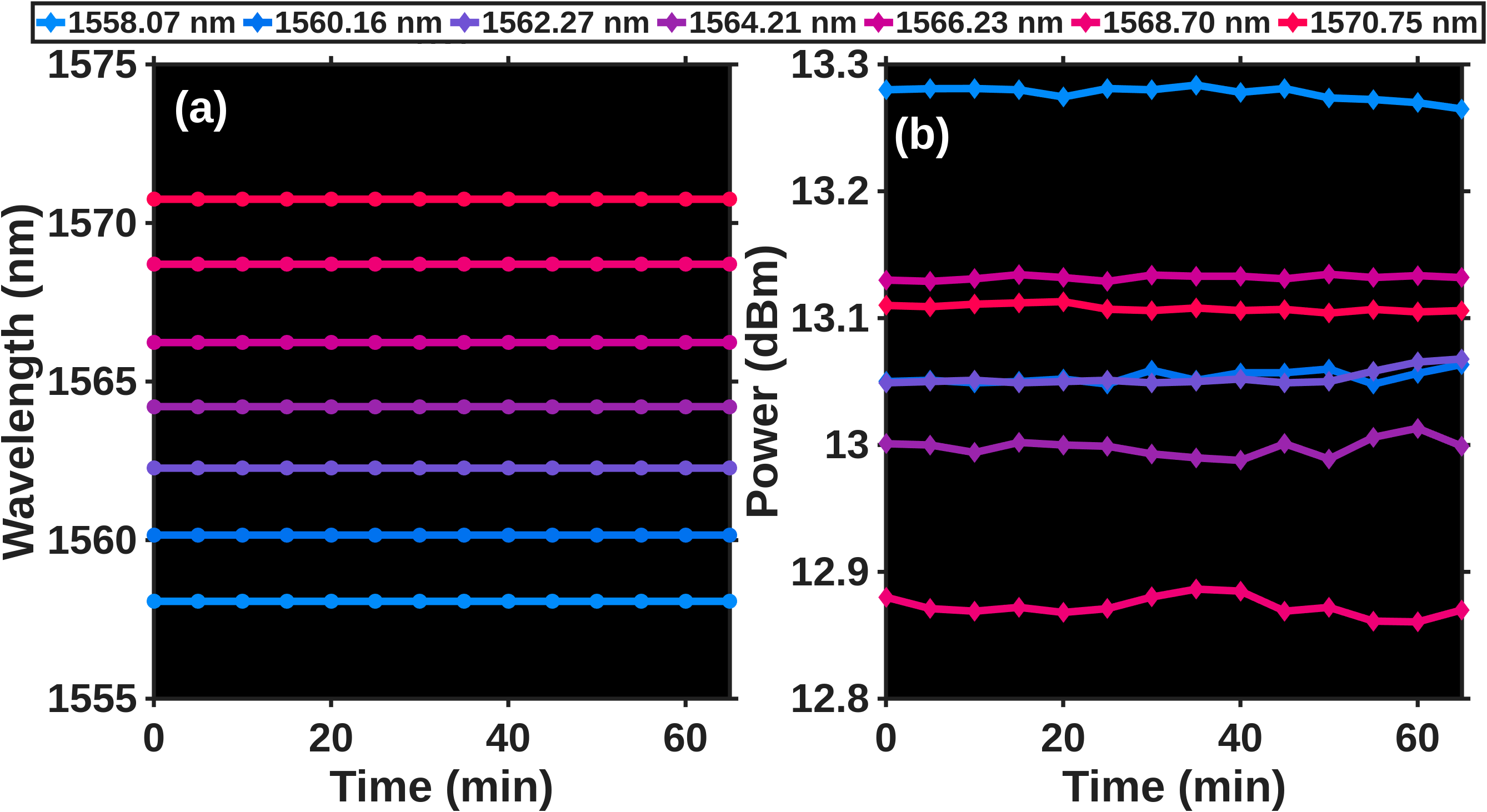} 
    \caption{Temporal stability of the seven-wavelength mode-locked laser, showing (a) wavelength drift and (b) power fluctuation over one hour operation time.}
    \label{stability}
\end{figure}
At a pump power of approximately 50 mW, stable seven-wavelength mode-locked operation is achieved. The corresponding optical spectrum, shown in Fig.~\ref{7 wl}, displays seven distinct and well-resolved lasing wavelengths extending across the C- and L-bands. The nearly constant spacing between adjacent spectral lines indicates that the multi-wavelength operation is governed by a periodic wavelength-selective transmission function imposed by the birefringence-induced filtering in the NPR cavity. Each lasing line exhibits a narrow spectral linewidth, with a typical 3-dB bandwidth between 0.10 to 0.17 nm, demonstrating strong spectral confinement and good phase coherence.

Notably, all seven wavelengths exhibit comparable output intensities. Furthermore, Kelly sidebands that are commonly observed in soliton mode-locked fiber lasers are either significantly suppressed or barely noticeable in the measured spectra. This behavior can be attributed to the effective intracavity spectral filtering induced by polarization evolution and fiber birefringence, which together form a Lyot-type filter within the cavity. By appropriately adjusting the intracavity polarization controllers, the extinction ratio of the Lyot filter can be modified. An increased extinction ratio results in a narrower effective filter bandwidth, which significantly affects the pulse dynamics. Under such strong filtering conditions, the pulse acquires an up-chirp and evolves toward a dissipative-soliton-like regime, where the combined action of spectral filtering and saturable absorption stabilizes the pulse~\cite{xu2022group}.

\begin{figure}[htb]
    \centering
    \includegraphics[trim={0.2cm 0.2cm 0.2cm 0cm},clip,,width=\linewidth]{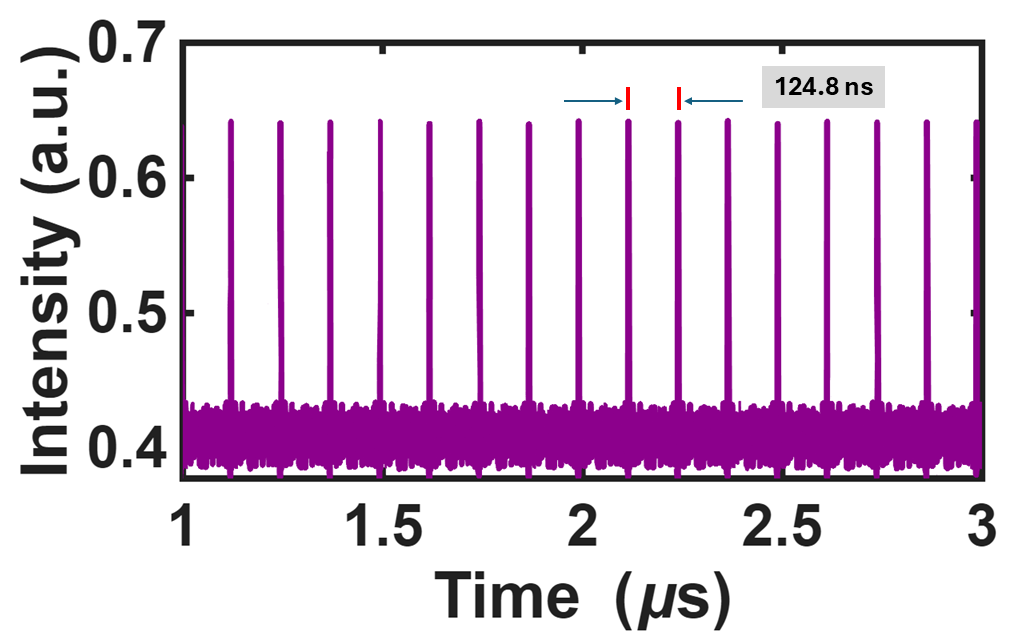} 
    \caption{Pulse train of seven-wavelength mode-locked state of the fiber laser.}
    \label{five}
\end{figure}

To evaluate the long-term stability of the seven-wavelength mode-locked operation, the time evolution of the emission wavelengths and their corresponding output powers was observed over one hour duration, as illustrated in Fig.~\ref{stability}. The set of traces depicted in Fig.~\ref{stability}(a) represents the drift of the seven lasing wavelengths, while Fig.~\ref{stability}(b) depicts the corresponding output power fluctuations as a function of time.
Evidently, all seven lasing wavelengths remain remarkably stable over the entire observation period, exhibiting negligible wavelength drift and effective suppression of mode hopping, confirmed by the straight horizontal lines in the wavelength traces (Fig.\ref{stability}(a)). 
Whereas only minor power fluctuations are observed for each wavelength peak, which remain within a narrow range throughout the measurement. These fluctuations are attributed primarily to 

environmental perturbations and intrinsic gain dynamics of the erbium-doped fiber, rather than instability of the mode-locking mechanism. Importantly, the absence of abrupt power variations or wavelength jumps confirms that the multi-wavelength emission remains phase-locked and dynamically stable.\\
The temporal characteristics of the seven-wavelength mode-locked state are illustrated in Fig.~\ref{five}. The recorded pulse train exhibits a uniform temporal separation of approximately 124.8 ns, which is identical to that observed in the single-wavelength operating regime and corresponds precisely to the cavity round-trip time. 



\begin{table*}[t]
\caption{Two-wavelength tunability in the same direction.}
\label{2wl_tune_t}
\centering

\begin{tabular}{ccccc}
\toprule
\textbf{Position} & \textbf{1st} & \textbf{2nd} & \textbf{3rd} & \textbf{4th} \\
\midrule
\textbf{Peaks (nm)} 
& 1560.54 / 1563.02 
& 1562.52 / 1565.21 
& 1565.83 / 1568.29 
& 1568.76 / 1571.50 \\
\addlinespace
\textbf{Separation (nm)} 
& 2.48 
& 2.69 
& 2.46 
& 2.74 \\
\bottomrule
\end{tabular}

\end{table*}

\begin{figure}[htb]
    \centering
\includegraphics[trim={0.2cm 0.2cm 0.2cm 0.2cm},clip,width=\linewidth]{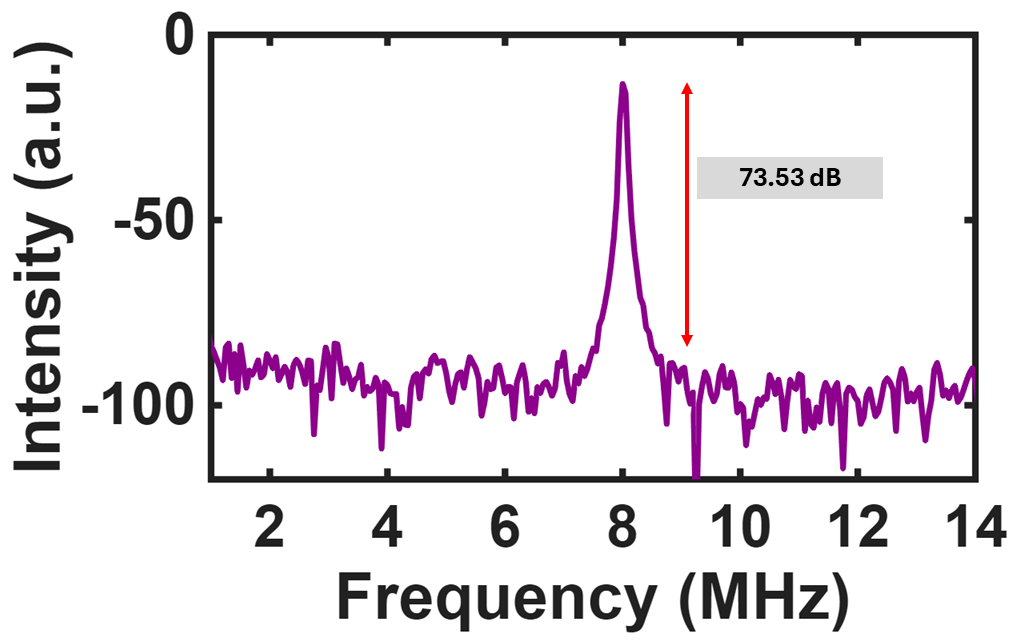} 
    \caption{RF spectra of the seven-wavelength mode-locked regime.}
    \label{rf 7}
\end{figure}

 These results confirm that all seven wavelengths remain phase-locked and temporally synchronized, sharing a common repetition frequency, demonstrating that pulses remain temporally overlapped and propagate as a group-velocity-locked state. Notably, this synchronization occurs naturally within the cavity without the need for additional pulse-shaping or synchronization components.
To further assess the stability of the mode-locked operation, the radio-frequency (RF) spectrum of the laser output was measured. The fundamental RF peak appears at 8.014 MHz, in excellent agreement with the cavity repetition rate. As shown in Fig.~\ref{rf 7}, the RF spectrum exhibits a signal-to-noise ratio exceeding 70 dB, indicating low timing jitter, negligible amplitude fluctuations, and highly stable phase-locked operation of the multi-wavelength mode-locked laser.
\subsection{Tunability of the Mode-locked fiber laser}
The laser demonstrates highly flexible spectral tunability across several distinct multi-wavelength regimes. In contrast to conventional tunable fiber lasers, where wavelength tuning is generally restricted to a single lasing line, the present system supports synchronous and collective wavelength tuning involving single-, dual-, triple-, and four-wavelength mode-locked states.

\begin{figure}[htb]
    \centering
    \includegraphics[width=\linewidth]{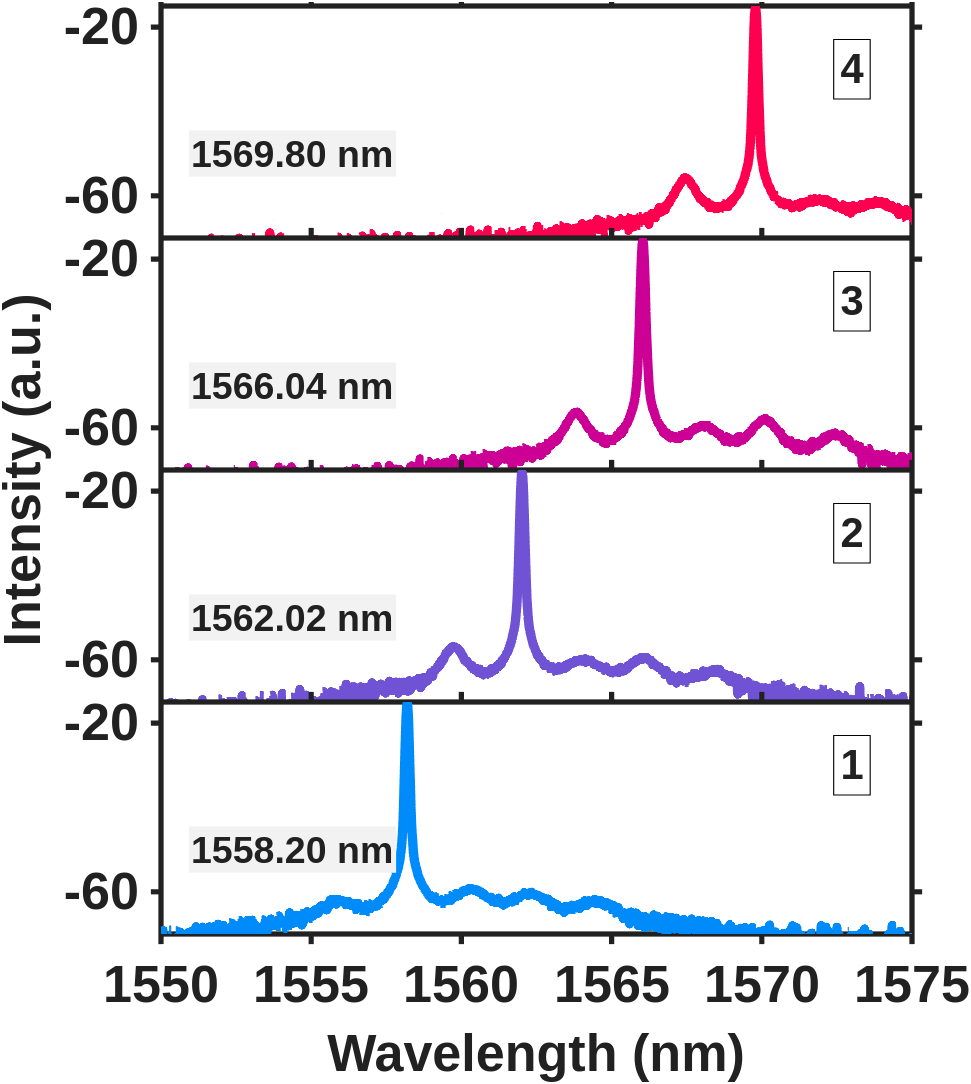} 
    \caption{Output spectra showing single-wavelength tuning over a range of 11.6 nm.}
    \label{tune1wl}
\end{figure}

\begin{figure}[htb]
    \centering
    \includegraphics[width=\linewidth]{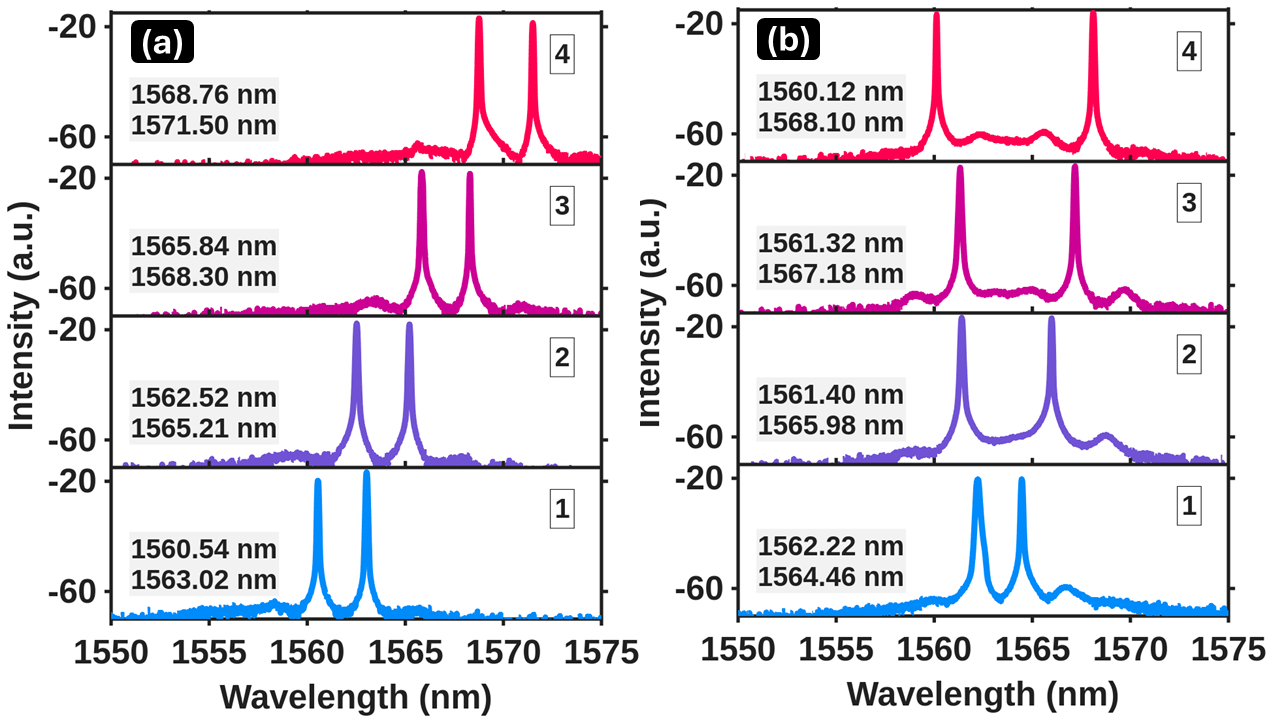} 
    \caption{Two wavelength tunability in the (a) same direction and (b) opposite direction.}
    \label{tune3}
\end{figure}

In the single-wavelength operating regime, continuous wavelength tunability is achieved by controlled rotation of the intracavity polarization controllers, which alter the effective cavity birefringence and shift the transmission maxima of the NPR-induced birefringent filter. This polarization-driven tuning mechanism enables smooth, reversible translation of the lasing wavelength over approximately 11.6 nm [Fig.~\ref{tune1wl}]. Fig.~\ref{tune3}(a) demonstrates dual-wavelength tunability, where two lasing peaks shift together as the polarization controllers are adjusted. Importantly, both wavelengths can be tuned simultaneously and synchronously up to 8.22 nm, while preserving a nearly constant spectral separation between the two peaks. In Table~\ref{2wl_tune_t}, the separation between the two wavelengths at different positions is mentioned.  This indicates that the dual-wavelength emission is governed by a common birefringent transmission comb, rather than independent gain competition. Under specific polarization configurations, asymmetric tuning behavior is observed, in which two wavelengths shift in opposite directions across the gain bandwidth, reaching a maximum deviation of up to 7.98 nm from their initial separation [Fig.~\ref{tune3}(b)]. This behavior reflects the polarization-dependent loss discrimination inherent to NPR, which allows selective control of individual spectral channels.

\begin{figure}[htb]
    \centering
    \includegraphics[width=\linewidth]{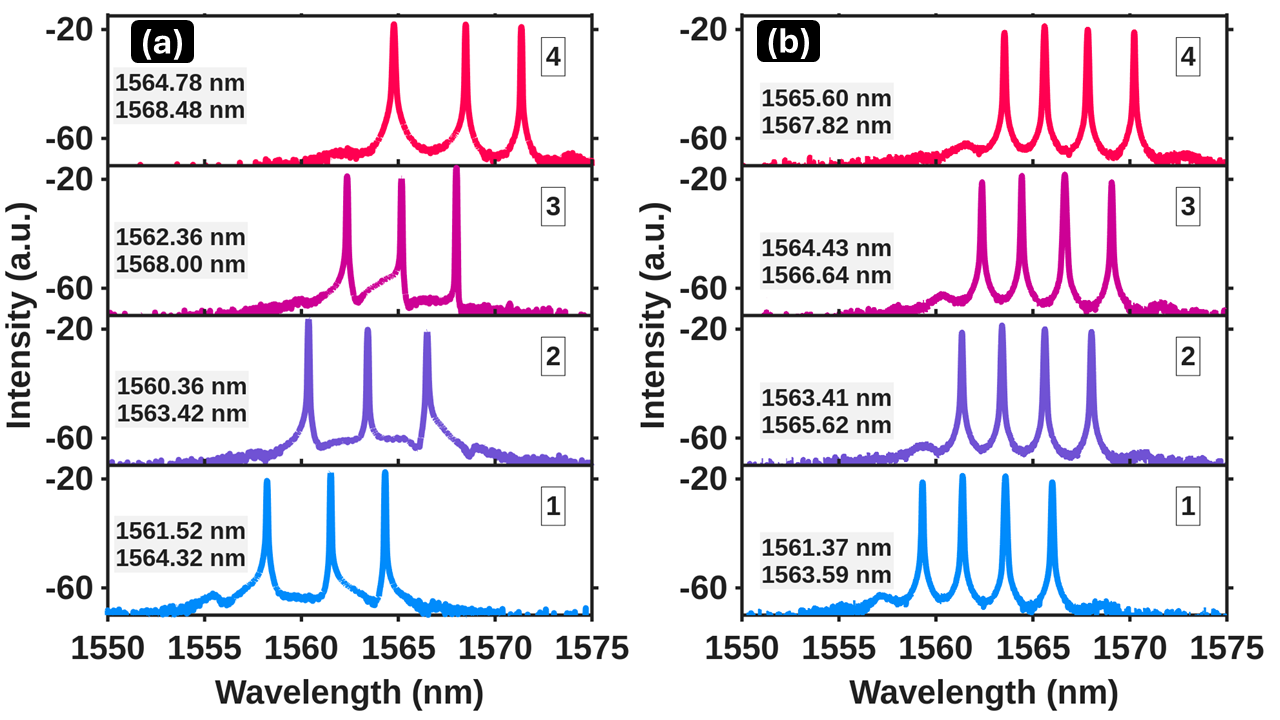} 
    \caption{(a) Three-wavelength, and (b) four-wavelength tunability.}
    \label{twl_3_4}
\end{figure}

Fig.~\ref{twl_3_4} further shows that the wavelength tunability observed in the single- and dual-wavelength regimes can be extended to higher-order multi-wavelength states, specifically the triple- and four-wavelength configurations, respectively, where several distinct lasing peaks are generated and translated collectively within the same cavity.

In the triple-wavelength regime, three well-resolved lasing channels shift simultaneously. A total tuning span of approximately 6.54 nm is achieved. Although slight variations in the inter-peak spacing are observed during the tuning process, the relative separation remains nearly constant across the tuning range. The peak separations are listed in Table~\ref{3wl_tunet}. The minor deviations can be attributed to the wavelength-dependent dispersion and birefringence of the fiber cavity, which introduce small differential phase shifts. Yet, such variations do not compromise the stability or coherence of the multi-wavelength emission.

In the four-wavelength regime, a similar collective spectral translation is observed. All four lasing peaks shift synchronously over a tuning range of approximately 4.23 nm under controlled polarization adjustment. In contrast to the triple-wavelength case, the inter-peak spacing remains essentially invariant throughout the tuning process, indicating stronger spectral locking among the channels. The negligible change in peak separation also confirms that the four wavelengths are governed by a common birefringence-induced transmission comb that translates quasi-rigidly across the erbium gain bandwidth. The physical origin of this collective tunability lies in the intensity- and wavelength-dependent loss modulation introduced by NPR. Rotation of the polarization controllers modifies the projection of the intracavity polarization state onto the analyzer axis of the polarization-dependent isolator, thereby reshaping the spectral loss landscape experienced by the circulating light.  As a result, a Lyot-type birefringent filter emerges, producing a periodic transmission spectrum with maxima separated by an FSR that is inversely proportional to the product of the cavity birefringence and fiber length. This comb defines the potential lasing windows. Moreover, the nonlinear Kerr effect introduces an intensity-dependent phase shift, which displaces the transmission comb relative to the erbium gain profile. This mechanism provides the fine tunability observed as wavelengths shift either collectively or symmetrically apart.

\begin{table*}[t]
\caption{Three-wavelength tunability in the same direction.}
\label{3wl_tunet}
\centering

\begin{tabular}{ccccc}
\toprule
\textbf{Position} & \textbf{1st} & \textbf{2nd} & \textbf{3rd} & \textbf{4th} \\
\midrule

\textbf{Peaks (nm)} 
& \makecell{1558.22 \\ 1561.50 \\ 1564.31}
& \makecell{1560.36 \\ 1563.44 \\ 1566.50}
& \makecell{1562.34 \\ 1565.18 \\ 1568.02}
& \makecell{1564.76 \\ 1568.50 \\ 1571.36} \\
\hline
\addlinespace

\textbf{First two peaks separation (nm)} 
& 3.28 & 3.08 & 2.84 & 3.74 \\

\textbf{Last two peaks separation (nm)} 
& 2.81 & 3.06 & 2.84 & 2.86 \\

\bottomrule
\end{tabular}

\end{table*}

\subsection{Switchable multi-wavelength ML fiber laser}
 
\begin{figure}[htb]
    \centering
    \includegraphics[width=\linewidth]{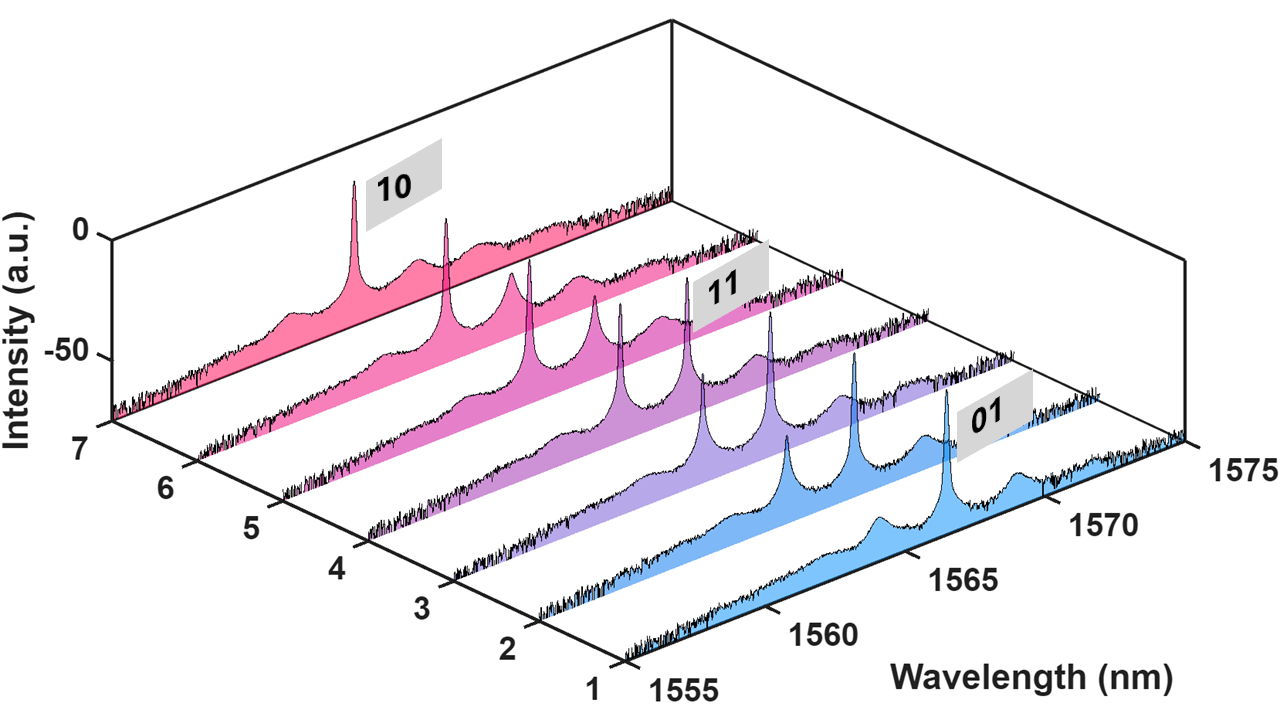} 
    \caption{Dynamic switching of a two-wavelength mode-locked fiber laser.}
    \label{sw 2wl 1}
\end{figure}

In addition to wavelength tunability, the laser cavity exhibits a highly versatile capability for spectral switchability. In this context, switchability refers to the ability of the system to realize abrupt, discrete, and fully reversible transitions between multiple well-defined and intrinsically stable spectral states. Unlike the gradual spectral evolution associated with continuous tuning, these transitions occur in a controlled, stepwise, and reproducible manner. Such behavior is of particular interest because the distinct spectral states can function as an optical binary system beneficial for optical signal processing and optical switching applications. As depicted in Fig.~\ref{sw 2wl 1}, dual-wavelength mode-locked operation is achieved with two well-defined lasing peaks centered at 1564.00 nm and 1566.38 nm. By precise adjustment of the intracavity polarization controllers, either wavelength can be selectively activated or suppressed without altering the pump power. This controllability enables the realization of a two-bit optical binary system in which the two wavelengths represent independent spectral states, allowing the cavity to access the full binary range of operation.\\
\begin{figure}[htb]
    \centering
    \includegraphics[width=\linewidth]{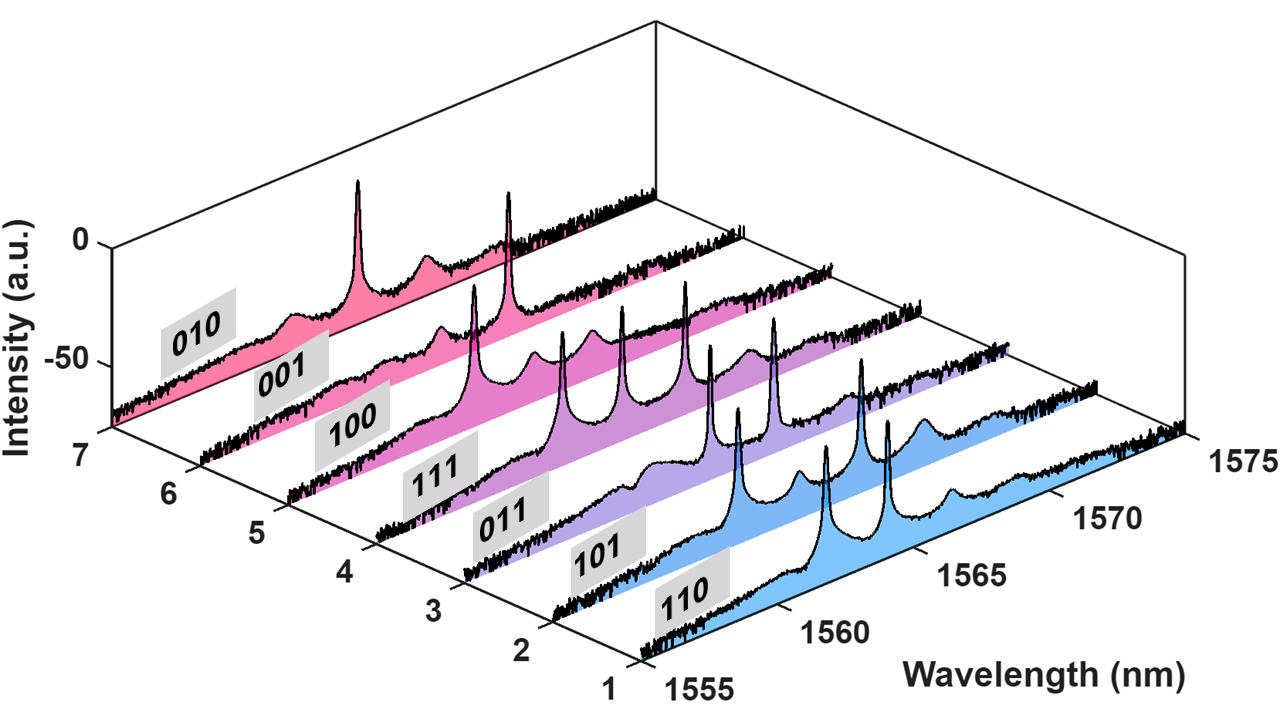} 
    \caption{Spectra of the three-wavelength mode-locked fiber laser demonstrating wavelength switchability across the entire three-bit binary operation.}
    \label{3wl switch}
\end{figure}
\begin{figure*}[htbp]
    \centering
    \includegraphics[width=\linewidth]{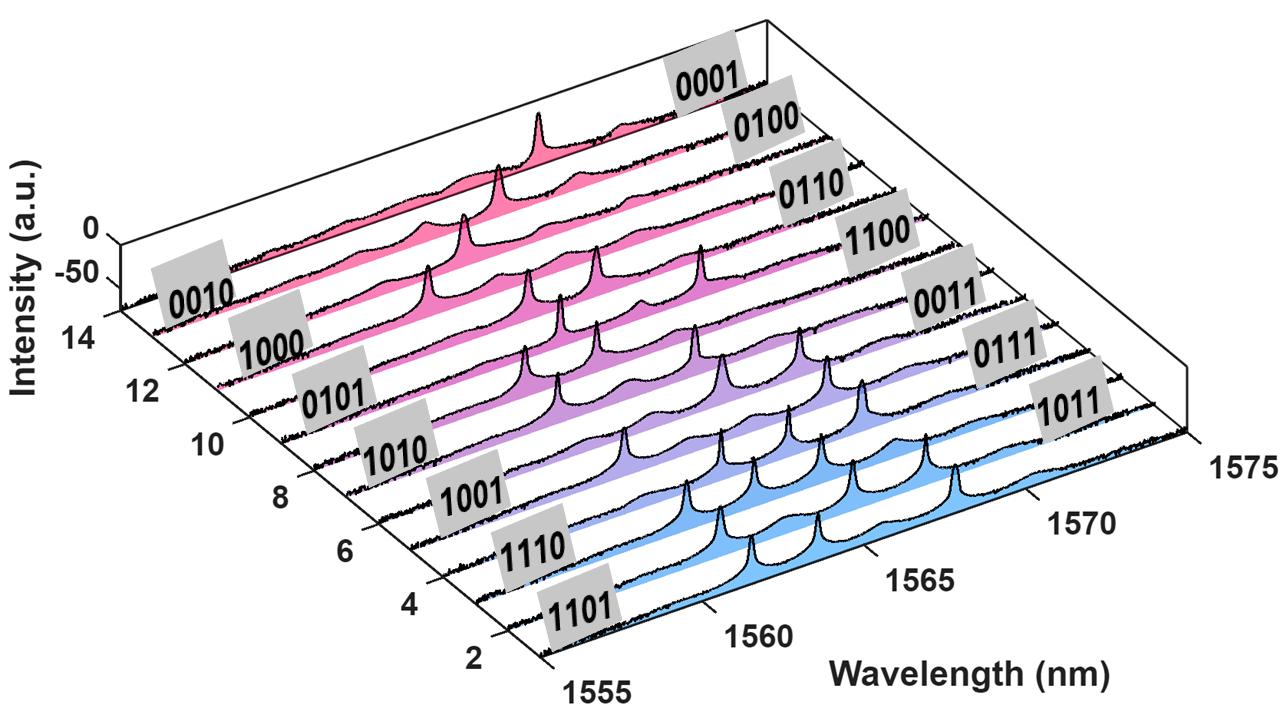} 
    \caption{Output spectra of the four-wavelength mode-locked fiber laser illustrating four-bit binary operation through selective suppression of wavelength peaks.}
    \label{4wl switch sup}
\end{figure*}
Figure~\ref{sw 2wl 1} illustrates the stepwise spectral switching dynamics under gradual polarization rotation. Initially, the cavity operates in a stable single-wavelength mode-locked regime centered at 1566.38 nm, with no additional spectral components present. As the polarization controllers are gradually rotated, a second lasing peak at 1564 nm emerges, leading to an intermediate regime in which both spectral components coexist. At this stage, stable dual-wavelength mode locking is established. With further polarization adjustment, the 1566.38 nm component progressively decreases in intensity, and the system ultimately evolves into a new stable single-wavelength state centered at 1564.00 nm, while the original 1566.38 nm peak becomes completely suppressed.

Figure~\ref {3wl switch} presents stable triple-wavelength mode-locked operation with three clearly resolved lasing peaks centered at 1561.84 nm, 1564.04 nm, and 1566.34 nm. The peaks exhibit well-defined spectral separation and comparable intensity levels, indicating effective birefringence-controlled filtering within the NPR-based cavity. By fine adjustment of the PCs, the three wavelength channels can be independently activated or suppressed in arbitrary combinations, enabling controlled transitions between single-, dual-, and triple-wavelength mode-locked regimes. In this configuration, each wavelength channel can be regarded as an independent binary state (bit), where the presence or suppression of a spectral peak corresponds to logical “1” or “0”, respectively. Consequently, the three lasing wavelengths constitute a three-bit optical binary system. Through gradual polarization adjustment, the cavity can evolve among different binary states. The simultaneous presence of all three wavelengths corresponds to the 111 state. Selective suppression of one wavelength leads to dual-wavelength states (110, 101, or 011), while further polarization tuning allows the cavity to transition to single-wavelength states (100, 010, or 001). These transitions occur without modifying the pump power or cavity configuration, indicating that the intracavity polarization state governs the effective wavelength-dependent loss in the cavity.\\

Furthermore, stable four-wavelength mode-locked operation is obtained with four distinct and well-resolved lasing peaks centered at 1561.52 nm, 1563.60 nm, 1565.68 nm, and 1567.92 nm, constituting a four-bit optical binary system. Similarly, by fine-tuning the PCs, arbitrary combinations of the four wavelength channels can be selectively activated or suppressed without altering the pump power or cavity configuration. Starting from the four-wavelength state (1111), selective suppression of one channel produces triple-wavelength states (1110, 1101, 1011, or 0111) [Fig.~\ref{4wl switch sup}]. Further polarization tuning enables dual-wavelength states (1100, 1010, 1001, 0110, 0101, or 0011), while simultaneous suppression of three channels results in single-wavelength operation (1000, 0100, 0010, or 0001). The switching process is abrupt, reversible, and repeatable, indicating that wavelength selection is governed by polarization-controlled, wavelength-dependent cavity loss induced by NPR rather than stochastic gain competition. 

These observations confirm that, in the wavelength-switchable regime, each lasing channel can be independently addressed and controlled without drift in its wavelength peak. The system allows the selective realization of any individual wavelength or any pairwise combination from the available channels within the same cavity architecture.  This full combinational controllability highlights the robustness of the NPR-induced birefringent filtering mechanism and provides an effective approach for polarization-controlled filter reconfiguration while maintaining stable mode-locking, resulting in highly versatile, reconfigurable, and deterministic multiwavelength operation within a fixed all-fiber cavity.

\section{conclusion}
In summary, we have demonstrated a fully all-fiber erbium-doped fiber laser passively mode-locked via NPR that simultaneously provides high-order multi-wavelength operation, collective tunability, and deterministic spectral switchability within a single fixed cavity. By controlling the intracavity birefringence and polarization state, the laser supports stable mode-locked operation from a single wavelength up to seven wavelengths, with narrow linewidths, uniform intensity distribution, and high RF signal-to-noise ratios.

Continuous wavelength tuning is achieved in the single-wavelength regime over a span of 11.6 nm, while in dual-, triple-, and four-wavelength regimes the lasing peaks translate synchronously with nearly constant inter-channel spacing. In addition, discrete and reversible spectral switching among single-, dual-, triple-, and quadruple-wavelength states, including arbitrary subset combinations, is realized solely through polarization control, without altering the pump power or cavity configuration. These behaviors arise from the interplay between birefringent comb filtering and Kerr-induced nonlinear phase modulation, leading to a periodic modulation of the net cavity gain inherent to the NPR mechanism.

The compact, alignment-free, and purely fiber-based architecture eliminates the need for external filters or modulators while offering exceptional spectral flexibility and temporal synchronization. These results establish NPR-based erbium-doped fiber lasers as highly reconfigurable ultrafast multi-wavelength sources with strong potential for applications in DWDM systems, fiber-optic sensing, spectroscopy, and advanced photonic signal processing.


%

~
~

~
~~

\section*{Acknowledgements}
The authors acknowledge CEFIPRA/IFCPAR (IFC/7148/2023) and UKIERI-SPARC (3673) for financial support through research projects. Additionally, KN thanks the Anusandhan National Research Foundation~(ANRF) for support through the Core Research Grant (CRG/2023/008068). Subrata Manna~(Ref- 2003349) and Amala Jose~(Ref- 2002215) gratefully acknowledge PMRF for the financial support through the PMRF fellowships.

\bibliography{Reference}
\end{document}